\newcommand{\be}{\begin{eqnarray}}
\newcommand{\ee}{\end{eqnarray}}

\documentstyle[galley,epsf]{mn}
\begin{document}
\title[$\beta$ CrB]{Radial Velocity Variations in Pulsating Ap Stars.\\
III. The Discovery of 16.21 min  Oscillations in $\beta$ CrB \footnote{Based on observations made at McDonald Observatory}}
\author[Hatzes \& Mkrtichian ]{
A.P. Hatzes$^1$ and D.E. Mkrtichian$^{2,3}$ \\
$^1$ Th\"uringer Landessternwarte Tautenburg, Sternwarte 5, D-07778, Tautenburg,
Germany\\
$^2$Astrophysical Research Center for the Structure and Evolution of the Cosmos, Sejong University, Seoul 143-747, Korea\\
$^3$Astronomical Observatory, Odessa National University, Shevchenko Park, Odessa, 65014, Ukraine\\
}
\date{}

\maketitle

\begin{abstract}
We present the analysis of  3 hrs of  a rapid time series
of precise stellar radial velocity (RV) measurements
($\sigma$ = 4.5 m\,s$^{-1}$) of the cool Ap star $\beta$ CrB.
The integrated RV measurements spanning the wavelength interval 5000-6000\,{\AA}
show significant variations (false alarm probability = 10$^{-5}$) with a period
of 16.21 min ($\nu$ = 1028.17 $\mu$Hz) and an amplitude of 3.54 $\pm$ 0.56 m\,s$^{-1}$. The RV measured
over a much narrower wavelength interval reveals one spectral feature
at $\lambda$6272.0 {\AA} pulsating with the same 16.21 min period and an
amplitude of 138 $\pm$ 23 m\,s$^{-1}$.
These observations establish $\beta$ CrB to be a low-amplitude
rapidly oscillating Ap star.
\end{abstract}


\section{Introduction}

Kurtz (1989) first suggested that the cool magnetic Ap star
$\beta$ CrB was a prime
candidate to be a rapidly oscillating Ap (roAp)
star since it had stellar properties similar to several known
oscillating Ap
stars (HR 1217, 33 Lib, and $\gamma$ Equ).
The issue of whether $\beta$ CrB is
a pulsating star  has important implications for the
excitation mechanism in roAp stars as well as the possible existence
of an instability strip for roAp stars.

	Several investigators have searched for pulsational variations in
$\beta$ CrB using photometric measurements with null results
(Heller \& Kramer 1988; Kreidl 1991). Recently,
Kochukhov et al. (2002, hereafter K02) presented radial velocity (RV)
measurements
for $\beta$ CrB spanning the wavelength interval 6105 -- 6190 {\AA}.
Measurements of most lines were constant to a level of 20--30 m\,s$^{-1}$;
however, intriguing evidence for RV variations were found for one spectral
feature, Fe I $\lambda$6165.4 {\AA}. This feature had an amplitude of
71 $\pm$ 11 m\,s$^{-1}$ and a period of 11.5 $\pm$ 0.5 min.
Before we can place $\beta$ CrB in the family of roAp stars
the K02 result must be confirmed since
the false alarm probability for the detection
was rather high (= 0.016).  Here we present our own precise radial
velocity measurements for $\beta$ CrB and show that it is indeed a rapidly
oscillating Ap star, but not with the period found by K02.

\section{Observations}

	Spectral observations of $\beta$ CrB were made for
3 consecutive hours on the night of 31 January 1998 (JD  = 2450844.903220) using
the ``2dcoude'' echelle spectrograph  of the Harlan J. Smith
2.7m telescope of McDonald Observatory (Tull et al. 1995).	
According to the ephemeris
of Kurtz (1989) this corresponds to magnetic phase 0.11.
One hundred fifty-five
observations were made with
exposure times
of 50 secs which yielded a typical signal-to-noise ratio of 150 per
pixel.  The description of the
instrumental setup can be found in previous papers (Kanaan \& Hatzes 1998;
Mkrtichian et al. 2003).
Briefly, the wavelength coverage was 4700 -- 6700 {\AA} at a resolving power
of 60,000. Precise stellar radial velocities were achieved
via an iodine absorption cell placed in the optical light path.

\section{Results}

\subsection{Integrated RV Measurements}

To achieve  the  maximum RV measurement precision we first used
the full spectral region covered by  the iodine
reference absorption lines to produce a ``mean'' RV value
for all spectral lines.   In doing so we  excluded the two
reddest spectral orders that covered the wavelength range
6100--6300\,{\AA}. Here the iodine absorption lines are weak and this
degrades the RV precision. 
To determine the relative RV
for $\beta$ CrB
each spectral order was divided into 60 chunks so that we could model
the spatial (and temporal) variations of the instrumental profile (IP)
which can introduce significant RV errors. In each chunk the RV shift
was calculated along with the IP using the IP reconstruction
procedure outlined in
Valenti et al. (1996).
The RV
measurements from all chunks were then combined weighted by the inverse
square of the RV standard deviation for each chunk. RV measurements of
an individual chunk that differed by more than 4$\sigma$ from the
mean were rejected in the calculation of the mean.

\begin{figure}
\epsfxsize=8.5truecm
\epsffile{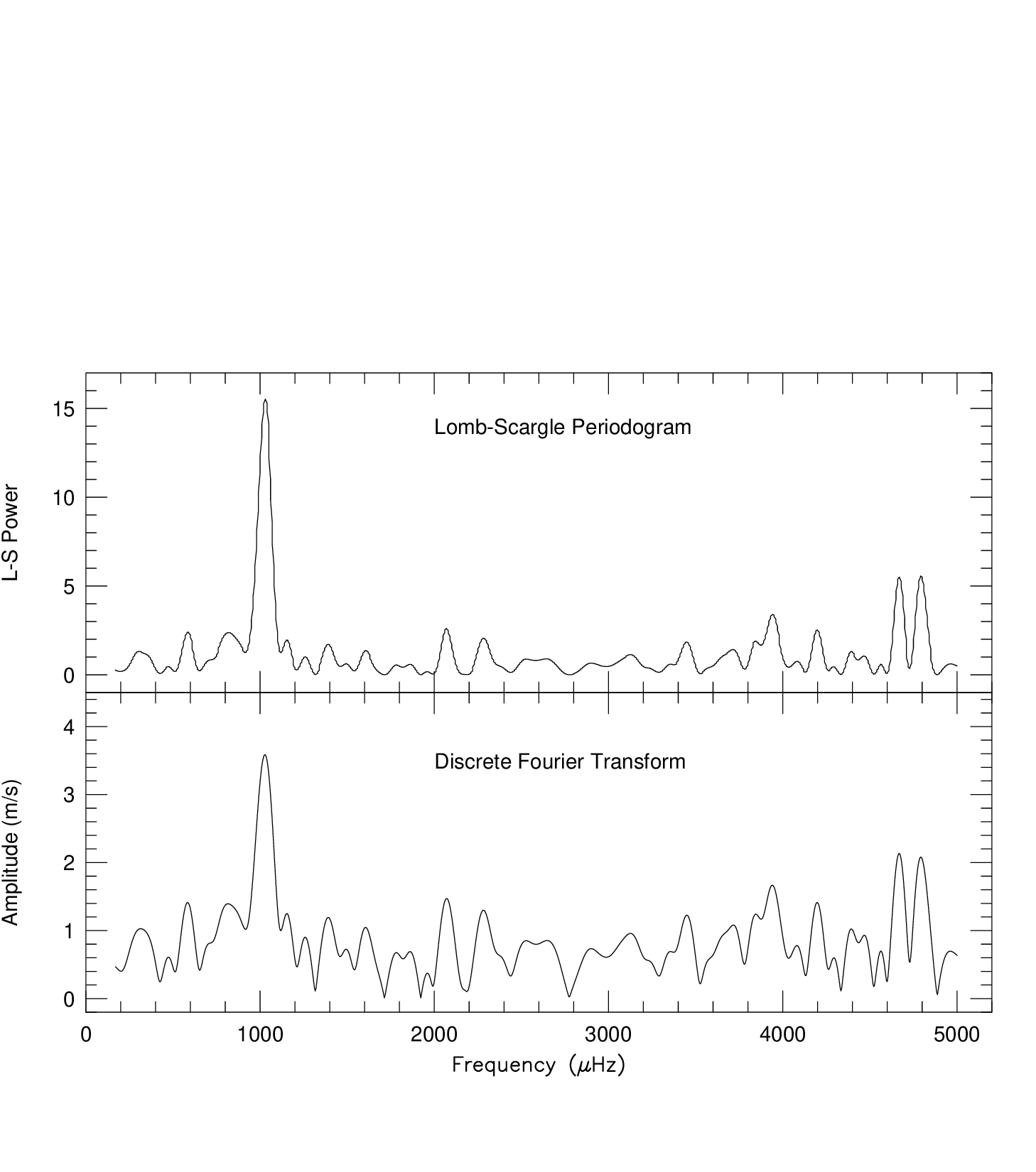}
\caption{(Top) The Lomb-Scargle (L-S) periodogram of the RV measurements
of $\beta$ CrB in the wavelength interval 5000--6000\,{\AA}.  The
L-S power (unitless)
is a measure of statistical significance. 
(Bottom) The discrete Fourier transform (amplitude spectrum)
of the same RV measurements. In this case amplitude is in units of
m\,s$^{-1}$}
\label{ft1} \end{figure}

The RV measurements  had  a 
long-term trend of unknown origin. This could be due to rotational
modulation caused  by the  abundance spots known to
occur on Ap stars, long-period oscillations, or a stellar or sub-stellar
companion. Observations with a longer timebase are needed to answer this.
Since we are primarily interested in searching for pulsational
variations on
time scales of roAp stars (5-15 min), a second order polynomial was fitted
to the long-term variations  and subtracted. Variations
in the sky transparency during the course of the observations resulted in some
spectra having significantly lower signal-to-noise ratio ($S/N$)  
than the mean ($S/N$ $\approx$ 150).
Observations having $S/N$ $<$ 100 (fifteen data points)
were eliminated from the time series analysis.

Figure~\ref{ft1} shows both the Lomb-Scargle (L-S) periodogram (Lomb
1976; Scargle 1982) and the discrete Fourier transform (DFT)
 of the residual RV time series for $\beta$ CrB. These show significant
power at 1028.17 $\mu$Hz ($P$ = 16.21 min).
When trying to detect
a periodic signal in time series data {\it and} assessing
its statistical significance the Lomb-Scargle (L-S) periodogram is a more
appropriate analysis tool over the DFT. In this case the L-S power is not a 
measure of the amplitude of the signal, but instead is related to 
the statistical significance of a peak. 
Many researchers in the field of oscillating stars are more
accustomed to dealing with DFT. In this instance the power
is directly related to the true amplitude of the periodic signal
in the data and can also be used to assess the noise level of the data.
The DFT, however, has limited use in quantifying the significance
of a signal. We thus show both the L-S periodogram and DFT (lower
panel of Figure~\ref{ft1}).

The L-S 
periodogram of the $\beta$ CrB data has power more  than 15 which is the
first hint that we are dealing with a statistically significant signal.
Using the Eq. 14 of Scargle (1982) this power results in  a false
alarm probability (FAP), i.e. the chance that noise is creating the signal,
of $\approx$ 2$\times$10$^{-5}$.

The power in a L-S periodogram only gives a rough estimate of the false
alarm probability of a signal. A more accurate determination is to 
use a bootstrap
randomization technique (e.g. K\"urster et al. 1997).
This technique
randomly shuffles the RV measurements keeping the times fixed.
A Lomb-Scargle periodogram 
is calculated for each ``random'' data set
and these data re-shuffled. After a large number of these  shuffles,
the
fraction of the periodograms having
maximum power greater than the observed data periodogram over the
frequency interval of interest (in this case $\nu$ = 
200--5000 $\mu$Hz) represents the
false alarm probability (FAP), or the chance that random noise can
produce the observed power in the periodogram.
Using 2$\times$10$^5$ shuffles we establish a FAP =
10$^{-5}$ for the peak in Figure 1.

\begin{figure}
\epsfxsize=8.5truecm
\epsffile{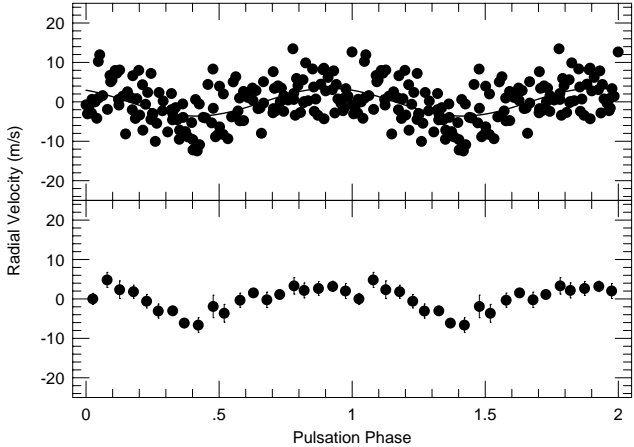}
\caption{(Top) The residual RV variations phased to a period
of 16.2 min. (Bottom) The phased-binned averages ($\Delta\phi$
$\approx$ 0.05) for data shown in the top panel. In both panels
data points are repeated for the second cycle.
}
\label{phase}
\end{figure}

Figure~\ref{phase} shows the residual RV measurements after removal
of the long-term trend phased to the 16.2 min period. The top panel is the
raw data while the bottom panel represents phase-binned ($\Delta\phi$
$\approx$ 0.05) values. The amplitude of these RV variations is
3.54 $\pm$ 0.56  m\,s$^{-1}$ ($K$-amplitude).

\begin{figure}
\epsfxsize=8.5truecm
\epsffile{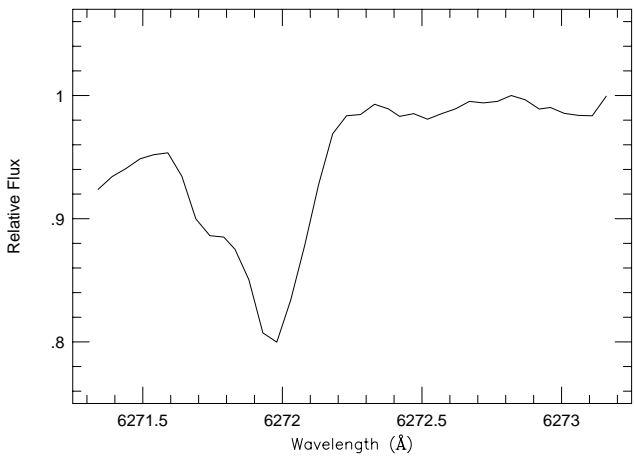}
\caption{Spectral region of the wavelength chunk ($\lambda$6271.7 -- $\lambda$6273.3\,{\AA})
showing high amplitude RV variations.
}
\label{spec}
\end{figure}

\begin{figure}
\epsfxsize=8.5truecm
\epsffile{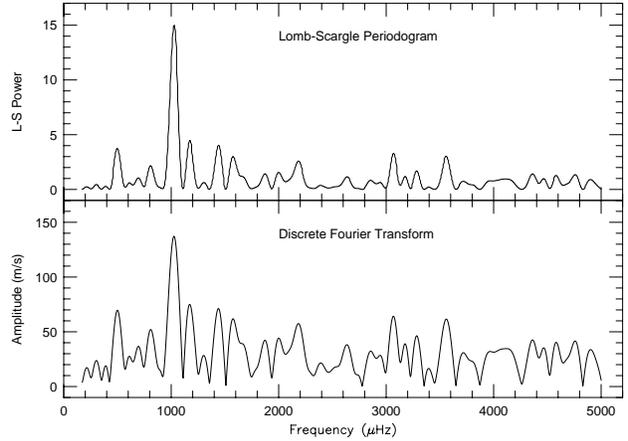}
\caption{The Lomb-Scargle periodogram (top) and DFT (bottom) of  the
RV variations for the spectral region  ($\lambda$6271.7 -- $\lambda$6273.3\,{\AA})
}
\label{ft2}
\end{figure}

\subsection{Narrow-band Measurements}

	As first shown by Kanaan \& Hatzes (1998) the RV amplitude
of individual spectral lines in roAp stars can differ
by factors of 10--100. Later studies established that
elements of Nd and Pr in  some roAp stars
had the highest pulsational RV amplitudes (Savanov et al. 1999,
Kochukhov \& Ryabchikova 2001; Kurtz et al. 2003).
To search for spectral lines that
that might have high amplitude RV variations the individual
wavelength chunks (each chunk typically
contained at most 1--2 spectral lines)
prior to
combining these in a mean velocity were examined.
Because high RV amplitude lines can also be found in the reddest
orders (excluded from the integrated RV analysis) these were also
used to search for RV variations).
A periodogram
analysis was performed on the RVs from the individual
wavelength chunks. If a periodogram showed significant power
(Lomb-Scargle power $>$ 7) a statistical assessment was made
using the bootstrap randomization technique (50,000--200,000 shuffles).

	Only one chunk ($\lambda$6271.7 -- $\lambda$6273.3\,{\AA})  had highly significant
variations with the same
period as found in the integrated RV data. The spectral region for
this chunk is shown in
Figure~\ref{spec}.
The RV amplitude ($K$) for this
chunk, 138 $\pm$ 23 m\,s$^{-1}$, was significantly higher than the
value for the integrated RV measurements.
Figure~\ref{ft2} shows both the Lomb-Scargle periodogram and DFT of the RVs
from this wavelength chunk. 
Using 200,000 shuffles we established  FAP $\approx$ 1.5 $\times$ 10$^{-5}$
for this peak.
The pulsational phase for this chunk is identical to within the errors of the
pulsation phase calculated using the integrated RV measurements.

	We tentatively identify the spectral features in the high
amplitude chunk as being a 
Ce\,II $\lambda$6272.026 {\AA} line that is blended
in the blue wing by a weaker Cr\,II $\lambda$6271.870\,{\AA} feature.

\section{Discussion}

\begin{figure}
\epsfxsize=8.5truecm
\epsffile{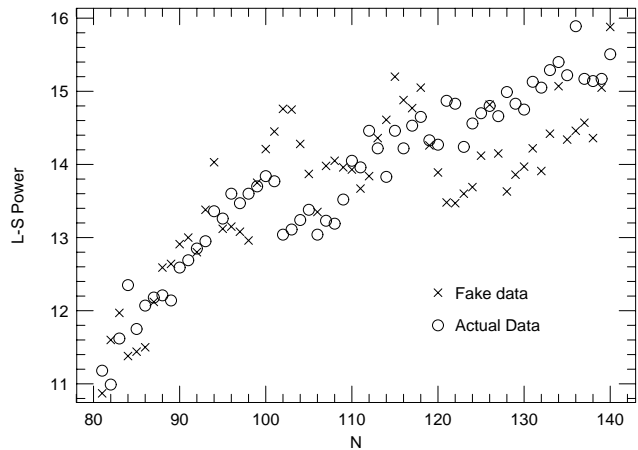}
\caption{The  Lomb-Scargle  power  at $\nu$ = 1028.17 $\mu$Hz
as a function of number of data points used in the periodogram using
real (circles) and fake data (crosses). The  fake data was generated
using a sine wave with the same amplitude (3.56 m\,s$^{-1}$)
and period (= 16.21 min) as found in our data and sampled in the same
manner. Random noise with the same scatter ($\sigma$ = 4.5
 m\,s$^{-1}$) as our data was also added
to the fake signal.
}
\label{powerinc}
\end{figure}

	Our precise RV measurements for $\beta$  CrB establish that
this star is with high probability a rapidly oscillating Ap star with an
amplitude
of 3.54 $\pm$ 0.56 m\,s$^{-1}$ and a period of 16.2 $\pm$ 0.7 min.
Although additional observations are needed to confirm our result,
we believe that the signal we have detected is real for a number of reasons:

\begin{itemize}
\item {\it We derive a very low false alarm probability.}

The probability that this is a false signal due to noise is
$\approx$ 10$^{-5}$ as determined through the bootstrap
randomization procedure. This is a more rigorous way of determining
the FAP than strictly using the Lomb-Scargle power.
This result represents the combined mean velocity of all spectral lines spanning
the wavelength interval 5000--6000\,{\AA}.  Any systematic or instrumental
error would have to affect all wavelength chunks used in the analysis in the
same way which seems unlikely.

\item {\it The variations are not due to a few outliers.}

	We think it is unlikely that a few outliers are causing our signal.
(We define an outlier as deviating significantly from the mean RV value
and not from the best fit sine wave through the data.) If only a few
outliers are driving the power in the periodogram then this would be evident
in a high false alarm probability.

For a real periodic signal in the presence of noise, the more data one
accumulates  the more significant the detection becomes, even with
the presence of a few occasional outliers.
Thus the power in a Lomb-Scargle periodogram should increase
with increasing number of data points.  This is demonstrated in 
Figure~\ref{powerinc} which shows the Lomb-Scargle power at 
$\nu$ = 1028.17 $\mu$Hz
(circles) as a
function of the number of data points used in the periodogram.  For comparison we generated
a synthetic signal consisting of a sine wave with the same period and 
amplitude as found in our data and sampled in the same manner
as the real data. Random noise was added with the same rms scatter as our
measurements. The crosses in Figure~\ref{powerinc}  show that the
 L-S power of the fake data (with a signal present)
increases with increasing number of data points in the
same way.

	An iodine absorption cell was used to perform our RV measurements
because it is designed to eliminate, or at least minimize instrumental effects.
However, one could argue that there might be residual instrumental shifts
not corrected by the reduction process, or that our instrumental profile
modelling
somehow introduces a false signal into the data. We do not believe this
is the case.

We investigated whether periodic instrumental shifts were present in our
data using the wavelength solution calculated
from the iodine absorption lines. A Fourier analysis of these shifts
showed the strongest period at 
176 minutes, and smaller variations with periods of 20 and 24 minutes.
All periods were significantly different from the one found in our RV
data. Furthermore, we also performed a Fourier analysis of
all IP parameters used in our RV determination. We found
no variations at the period coinciding with the one found in the RV
data. This demonstrates that using the iodine cell technique 
does an excellent job of excluding instrumental
variations. 

\item {\it The same signal is found in the $\lambda$6272 {\AA} feature.}

	The spectral orders used to determine the integrated RV 
did not include the order covering the $\lambda$6272 {\AA}
feature. Thus this spectral line is not contributing to the signal
found in our integrated RV measurements. However, a separate
analysis of this  spectral chunk did find the same 16.21 min period
and with the same statistical significance as with the integrated RV
measurements. Although this is not an independent confirmation
(we are using the same data set, but different parts of the spectrum),
this result and the above arguments argue strongly that the
16.21 min oscillations in $\beta$ CrB are real.

\end{itemize}

        Our RV measurements fail to confirm the results of K02
on the Fe I $\lambda$6165\,{\AA} feature.
The wavelength chunk containing this spectral line had an
rms scatter
of 130 m\,s$^{-1}$ (due to the weak iodine absorption lines
in this wavelength region). However, we had a factor
of 3 more measurements than K02. Monte Carlo simulations
indicate that we would	have detected an 11.5 min period with an amplitude
of 70 m\,s$^{-1}$ and a
FAP of 10$^{-4}$. An amplitude of 50 m\,s$^{-1}$  would have been
detected with a FAP = 0.01. Thus we would have detected any 
variations in the Fe I $\lambda$6165\,{\AA} line if they were present at the
same amplitude as reported by K02.

We believe that the result of K02 on $\beta$ CrB is spurious. The
FAP is much too high (FAP = 0.016) and our measurements have failed
to find the presence of an 11.5 min period in either the integrated RV
measurements, or in the analysis of the narrow wavelength chunks.
More troubling is that the K02 period of 11.5 $\pm$0.5 min found
in $\beta$\,CrB is very close to the values of periods
($\approx$ 11.7 min) found in several spectral lines in the roAp star
10 Aql using the same instrument. 

The most likely explanation for the RV signal found in $\beta$ CrB
by K02 is that it is due to short term uncorrected instrumental variations
in the Gecko spectrograph  used for the measurements.
Circumstantial evidence for this comes from 
Matthews \& Scott (1995) who reported an 
11.1 min  period in the RV variations in  $\gamma$\,Equ. This period is 
different from known pulsation modes and has never been confirmed
by subsequent RV measurements from other investigations. Although
the authors used  a superimposed mercury emission line from an arc lamp
to eliminate instrumental shifts, the 11.1 min period could still be a
residual instrumental effect due to the stellar light a calibration source
having a slightly different optical path (not the case when using
the iodine absorption cell).

The RV measurements of K02, on the other hand,
were made {\it without} a simultaneous
wavelength calibration (either I$_2$ absorption cell or simultaneous
Th-Ar calibration), rather the wavelength calibration used a Th-Ar
exposure taken before and after the time series used for the RV
measurements. This cannot correct for any instrumental variations on time
scales significantly shorter than the time between the two calibration
measurements. Because K02 have failed to exclude an instrumental origin
with a period of $\approx$ 11.5 min for the RV variations in $\beta$ CrB
their results on this star (and possibly 10 Aql) should be considered suspect. 
An investigation of the short-term instrumental shifts of the Gecko
spectrograph would be useful. 

	Our RV measurements
for $\beta$ CrB seem to establish that this star is
 an roAp star with the lowest RV amplitude
(3.5 m\,s$^{-1})$ and one of the longest periods (16.2 min).  $\beta$ CrB is also unique
among roAp stars in that it shows high amplitude pulsational in only one
spectral feature, the $\lambda$6271.9\,{\AA}. This showed
significant variations (FAP $=$ 1.5$\times$10$^{-5}
$)
with the same 16.2 min period and an amplitude of 138 $\pm$ 23
m\,s$^{-1}$.
We tentatively identify this feature as a  blend of Ce\,II and
Cr\,II lines.
Why $\beta$ CrB has such a low RV amplitude, almost no high amplitude
spectral lines,  and a
longer period mode compared to other roAp stars with similar spectral
properties should provide clues as to the origin of the roAp
phenomenon.
We are currently planning observations of this star over
a full rotation period not only to confirm our detection, but to search
for rotationally modulated amplitude variations of the pulsations.

\section{acknowledgements}
APH acknowledges the support of grant 50OW0204 from the
Deutsches Zentrum f\"ur Luft- und Raumfahrt e.V. (DLR).
DEM acknowledges his work as part of research activity of the
Astrophysical Research Center for the Structure and Evolution of the
Cosmos (ARCSEC) which is supported by the Korean Science and Engineering
Foundation. We thank William Cochran for useful discussions and for
his assistance in developing the radial velocity reduction software
and Antonio Kanaan for his help in acquiring the data.

{} 

\end{document}